# 環境適応ソフトウェアにおけるアプリケーション最適配置

山登庸次†

† NTT ネットワークサービスシステム研究所，東京都武蔵野市緑町 3-9-11
E-mail: †yoji.yamato.wa@hco.ntt.co.jp

**あらまし** ヘテロハードウェアを活用するには，プログラマーは OpenMP，CUDA，OpenCL 等の十分な技術スキルが必要であった．そこで，私は，一度記述したコードの自動変換，構成を行い，高性能操作を可能にする環境適応ソフトウェアを提案し，自動変換等に取り組んできた．しかし，ユーザーの価格と応答時間の要件を満たすために変換されたアプリケーションをどこに配置するかについての検討はされていなかった．本論文では，環境適応ソフトウェアの新しい要素として，線形計画法を用いて変換アプリケーションの適切な配置を計算する方法を検討する．ユーザ要望等の条件を変化させたシミュレーション実験を通じてアプリケーションを適切に配置できることを確認した．

# Application placement study of environment adaptive software

Yoji YAMATO†

† Network Service Systems Laboratories, NTT Corporation, 3-9-11, Midori-cho, Musashino-shi, Tokyo
E-mail: †yoji.yamato.wa@hco.ntt.co.jp

**Abstract** To use heterogeneous hardware, programmers must have sufficient technical skills to utilize OpenMP, CUDA, and OpenCL. On the basis of this, I have proposed environment-adaptive software that enables automatic conversion, configuration, and high performance operation of once written code, in accordance with the hardware. However, although it has been considered to convert the code according to the offload devices, there has been no study where to place the offloaded applications to satisfy users' requirements of price and response time. In this paper, as a new element of environment-adapted software, I examine a method to calculate appropriate locations using linear programming method. I confirm that applications can be arranged appropriately through simulations.

## 1. はじめに

近年，CPU の半導体集積度が 1.5 年で 2 倍になるというムーアの法則が減速するのではないかと言われている．そのような状況から，CPU だけでなく，FPGA（Field Programmable Gate Array）や GPU（Graphics Processing Unit）等のデバイスの活用が増えている．例えば，Microsoft 社は FPGA を使って Bing の検索効率を高めるといった取り組みをしており [1]，Amazon 社は，FPGA, GPU 等をクラウド技術を用いて（例えば, [2]- [8]）インスタンスとして提供している [9]．また，システムでは IoT 機器利用（例えば, [10]- [19]) 等も増えている．

しかし，少コアの CPU 以外のデバイスをシステムで適切に活用するためには，デバイス特性を意識した設定やプログラム作成が必要であり，OpenMP（Open Multi-Processing）[20]，OpenCL（Open Computing Language）[21]，CUDA（Compute Unified Device Architecture）[22] といった知識や IoT 機器向けの組み込みシステムの知識が必要になってくるため，大半のプログラマーにとっては，スキルの壁が高い．

少コアの CPU 以外の GPU や FPGA，メニーコア CPU 等のデバイスを活用するシステムは今後ますます増えていくと予想されるが，それらを最大限活用するには，技術的壁が高い．そこで，そのような壁を取り払い，少コアの CPU 以外のデバイスを十分利用できるようにするため，プログラマーが処理ロジックを記述したソフトウェアを，配置先の環境（FPGA，GPU，メニーコア CPU 等）にあわせて，適応的に変換，配置し，環境に適合した動作をさせるような，プラットフォームが求められている．

Java [23] は 1995 年に登場し，一度記述したコードを，別メーカーの CPU を備える機器でも動作可能にし，環境適応に関するパラダイムシフトをソフト開発現場に起こした．しかし，移行先での性能については，適切であるとは限らなかった．そこで，私は，一度記述したコードを，配置先の環境に存在する

GPU や FPGA，メニーコア CPU 等を利用できるように，変換，リソース設定等を自動で行い，アプリケーションを高性能に動作させることを目的とした，環境適応ソフトウェアを提案した．合わせて，環境適応ソフトウェアの要素として，アプリケーションコードのループ文及び機能ブロックを，FPGA，GPU に自動オフロードする方式を提案評価している [24]- [28].

本稿は，通常の CPU 向けプログラムを，GPU 等のデバイスにオフロードした際に，アプリケーションをユーザのコスト等要求を満たして，応答時間等を短く動作するように，配置先を適正化するための手法を線形計画法を用いて提案する．アプリケーションタイプ，コスト要求等幾つかの条件を変えた際の，シミュレーション実験を通じて適切に配置できることを確認する．

## 2. 既 存 技 術

### 2.1 市 中 技 術

環境適応ソフトウェアとしては，Java がある．Java は，仮想実行環境である Java Virtual Machine により，一度記述した Java コードを再度のコンパイル不要で，異なるメーカー，異なる OS の CPU マシンで動作させている（Write Once, Run Anywhere）．しかしながら，移行先で，どの程度性能が出るかはわからず，移行先でのデバッグや性能に関するチューニングの稼働が大きい課題があった（Write Once, Debug Everywhere）．

GPU の並列計算パワーを画像処理でないものにも使う GPGPU（General Purpose GPU）（例えば [29]）を行うための環境として CUDA が普及している．CUDA は GPGPU 向けの NVIDIA 社の環境だが，FPGA，メニーコア CPU，GPU 等のヘテロなデバイスを同じように扱うための仕様として OpenCL が出ており，その開発環境 [30] も出てきている．CUDA，OpenCL は，C 言語の拡張を行いプログラムを行う形だが，プログラムの難度は高い（FPGA 等のカーネルと CPU のホストとの間のメモリデータのコピーや解放の記述を明示的に行う等）

CUDA や OpenCL に比べて，より簡易にヘテロなデバイスを利用するため，指示行ベースで，並列処理等を行う箇所を指定して，指示行に従ってコンパイラが，GPU，メニーコア CPU 等に向けて実行ファイルを作成する技術がある．仕様としては，OpenACC [31] や OpenMP 等，コンパイラとして PGI コンパイラ [32] や gcc 等がある．

CUDA，OpenCL，OpenACC，OpenMP 等の技術仕様を用いることで，FPGA や GPU，メニーコア CPU へオフロードすることは可能になっている．しかしデバイス処理自体は行えるようになっても，高速化することには課題がある．例えば，マルチコア CPU 向けに自動並列化機能を持つコンパイラとして，Intel コンパイラ [33] 等がある．これらは，自動並列化時に，コードの中のループ文中で並列処理可能な部分を抽出して，並列化している．しかし，メモリ処理等の影響で単に並列化可能ループ文を並列化しても性能がでないことも多い．FPGA や GPU 等で高速化する際には，OpenCL や CUDA の技術者がチューニングを繰り返したり，OpenACC コンパイラ等を用いて適切な並列処理範囲を探索し試行することがされている．

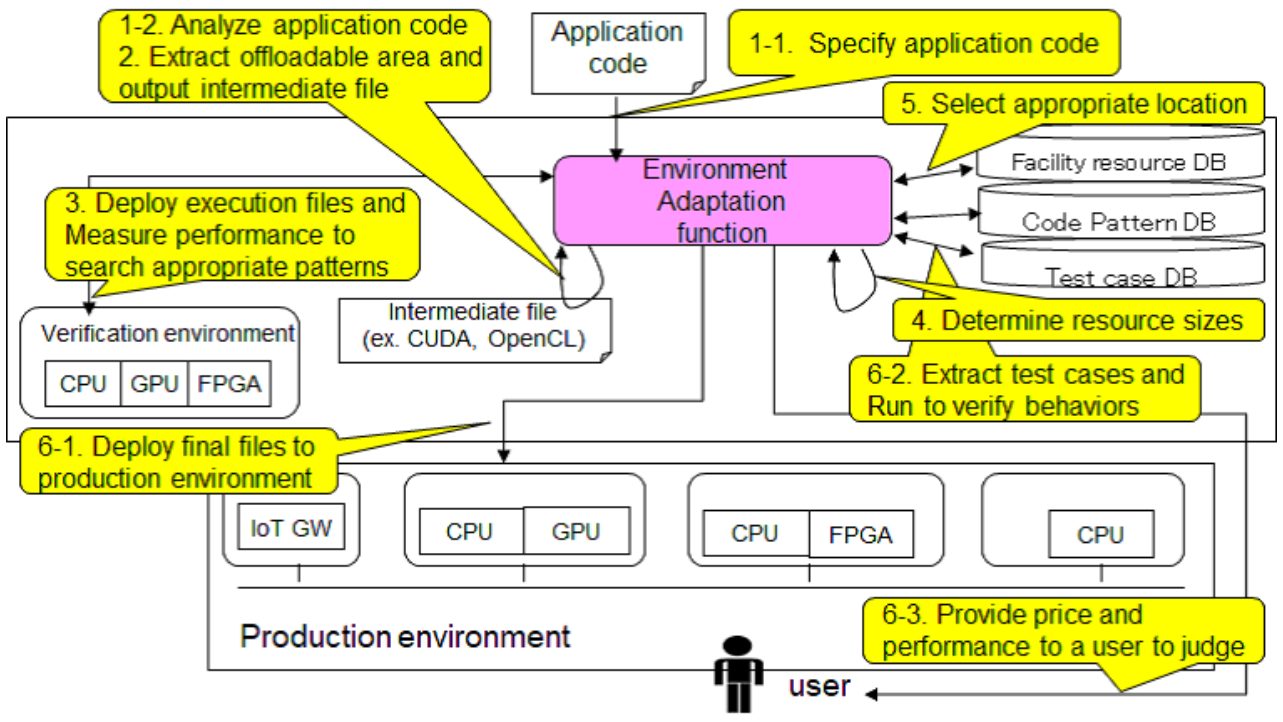

図 1 環境適応ソフトウェアのフロー

配置に関して，ネットワークリソースの最適利用として，ネットワーク上にあるサーバ 群に対して VN（Virtual Network）の埋め込み位置を最適化する研究がある [34]. [34] では，通信トラヒックを考慮した VN の最適配置を決定する．しかし，単一リソースの仮想ネットワークが対象で，キャリアの設備コストや全体的応答時間の削減が目的で，個々に異なるアプリケーションの処理時間や，個々のユーザのコストや応答時間要求等の条件は考慮されていない．

### 2.2 環境適応処理のフロー

ソフトウェアの環境適応を実現するため，著者は図 1 の処理フローを提案している．環境適応ソフトウェアは，環境適応機能を中心に，検証環境，商用環境，テストケース DB，コードパターン DB，設備リソース DB の機能群が連携することで動作する．

Step1 コード分析：

Step2 オフロード可能部抽出：

Step3 適切なオフロード部探索：

Step4 リソース量調整：

Step5 配置場所調整：

Step6 実行ファイル配置と動作検証：

Step7 運用中再構成：

ここで，Step 1-7 で，環境適応するために必要となる，コードの変換，リソース量の調整，配置場所の決定，検証，運用中の再構成を行うことができる．

現状を整理する．ヘテロなデバイスに対するオフロードは手動での取組みが主流である．著者は環境適応ソフトウェアのコンセプトを提案し，自動オフロードを検討しているが，自動オフロードした後の，Step5 にあたるアプリケーションの適切な配置については，検討がされていない．そのため，本稿では，GPU 等のオフロードデバイスに配置できるよう自動変換した際に，ユーザのコスト等の要求を満たして応答時間等を低減する配置手法を対象とする．

## 3. アプリケーション配置場所の適切化

### 3.1 配置考慮事項

[28] の手法では遺伝的アルゴリズム [35] を用いて，Clang [36] 等を用いて解析した通常のプログラムを，GPU 等のオフロー

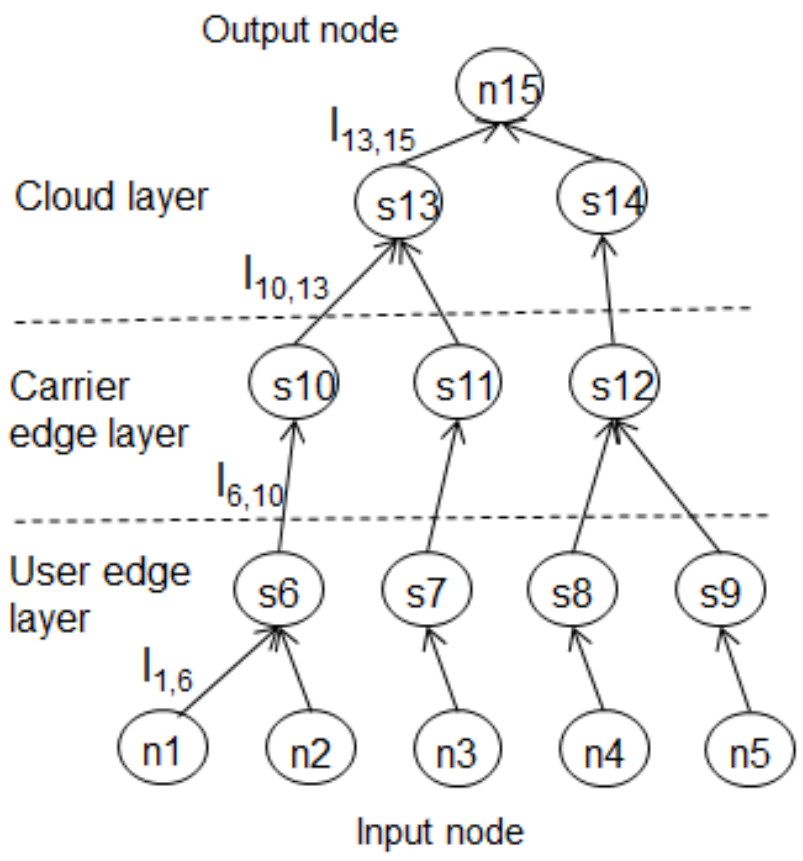

図 2　計算ノードのトポロジー例

ドデバイスに自動変換することができる．本サブ節では，デバイスにオフロードするプログラム変換が出来た後に，適切な場所へのアプリケーション配置を検討する．

[28] の手法では，複数のオフロードパターンを検証環境で繰り返し試行し，適切なオフロードパターンを選択している．そのため，変換されるアプリケーションは，最終的に選択されるオフロードパターン含め，各オフロードパターンでの処理時間，データ量，利用帯域，計算リソース量等が GPU 等にオフロードした場合で計測されている．各アプリケーションをオフロードした際のそれらの値は，適正配置計算で利用される．

従来，アプリケーションはクラウドに配置され，IoT デバイス等で収集したデータはクラウドサーバに転送され，そこでデータを集約し分析がされていた．しかし，現在エッジコンピューティングやフォグコンピューティングというキーワードで，リアルタイム性が必要な処理は，ユーザ環境やネットワークエッジで処理し，クラウドと処理分担することで，アプリケーションの品質を上げる試みが増えている．

本稿でも，アプリケーションはクラウドだけでなく，ネットワークエッジやユーザエッジに配置できる前提で検討する．ただし，ネットワークエッジやユーザエッジは，クラウドに比べサーバの集約度が低く分散しているため，計算リソースのコストはクラウドに比べ割高となる（一般に CPU や GPU 等のハードウェアの価格は配置場所によらず一定だが，クラウドを運用するデータセンタでは集約されたサーバをまとめて監視や空調制御等できるため，運用費が割安となる）．

例えば，計算ノードリンクの簡単なトポロジーとしては，図 2 等が考えられる．図 2 は，IoT システムのように，ユーザ環境でデータを収集する IoT デバイス等から，ユーザエッジにデータが送られ，ネットワークエッジを介してクラウドにデータが送られ，分析結果を会社の幹部が見る等で使われるトポロジーである．計算ノードは CPU，GPU，FPGA の 3 種に分けられる．GPU や FPGA を備えるノードには CPU も搭載されているが，仮想化技術（例えば，NVIDIA vGPU [37]）により，GPU インスタンス，FPGA インスタンスとして，CPU リソースも含む形で分割して提供される．

アプリケーションは，クラウド，ネットワークエッジ，ユーザエッジに配置され，ユーザ環境に近い側程，応答時間を低減する事が可能になる代わりに，計算リソースのコストが高くなる．本稿では，GPU や FPGA 向けに変換したアプリケーションを配置することになるが，配置する際に，ユーザは 2 種類のリクエストを発出できる．一つ目は，コスト要求であり，アプリケーションを動作させるために許容できるコストを指定する形で，例えば月 5000 円以内で動作させる等である．二つ目は，応答時間要求であり，アプリケーションを動作させる際の許容応答時間を指定する形で，例えば 10 秒以内に応答を返す等である．

従来から行われている設備設計では，[34] 等でも検討されているように，例えば仮想ネットワークを収容するサーバを配置する場所を，トラフィック増加量等の長期的傾向を見て，計画的に設計している．

それに対し，本稿では 2 つ特徴がある．一つ目は，配置されるアプリケーションは静的に定まっているのではなく，GPU や FPGA 向けに自動変換され，GA 等を通じて利用形態に適したパターンが実測を通じて抽出されるため，アプリケーションのコードや性能は動的に変わり得る（例えば，同じフーリエ変換プログラムでも，ユーザ A はサイズ大きいデータ，ユーザ B はサイズ小さいデータで使う場合に，GPU にオフロードされるループ文が異なったり，A 向けには CPU の 10 倍性能となり B 向けには 5 倍性能だったりする）．二つ目は，キャリアの設備コストや全体的応答時間だけを低減すれば良いのでなく，コストや応答時間に対する個々のユーザ要求を満たす必要があり，アプリケーションの配置ポリシーも動的に変わり得る．

この二つの特徴も踏まえ，本稿のアプリケーション配置は，ユーザからの配置依頼があったら，変換を行い，変換したアプリケーションをその時点で適切なサーバに順次配置していく形とする．アプリケーションを変換しても，コストパフォーマンスが向上しない場合は変換前を配置する．例えば，GPU インスタンスは CPU インスタンスの 2 倍のコストがかかる際に，変換しても 2 倍以上性能が改善されないならば，変換前を配置した方が良い．また，既に上限まで計算リソースや帯域が使われてしまっている場合はそのサーバには配置はできない．

### 3.2　アプリケーション適切配置のための線形計画式

本サブ節では，アプリケーションの適切な配置場所を計算するための，線形計画手法の定式化を行う．パラメータは図 3 に記載する．

ここで，デバイスやリンクのコストや計算リソース上限，帯域上限等は，事業者が準備するサーバやネットワークに依存するため，それらのパラメータ値は事業者が事前に設定する．オフロードした際にアプリケーションが使用する計算リソース量，帯域，データ容量，処理時間は，自動変換する前の検証環境での試験での最終的に選択されたオフロードパターンでの計測値により決まり，環境適応機能により自動設定される．

ユーザ要求がコスト要求か応答時間要求かで，目的関数と制約条件が変わる．コスト要求により，一月幾ら以内での配置が必要な要求の場合は，(1) の応答時間の最小化が目的関数とな

$a_i$: Device usage cost
$b_j$: Link usage cost
$C_i^d$: Device calculation resource limit of #i
$C_j^l$: Link bandwidth limit of #j
$C_k$: Data size of #k application
$A_{i,k}^d$: Whether to use of #k application on #i device
$A_{j,k}^l$: Whether to use of #k application on #j link
$B_k^d$: Calculation resource of #k application
$B_k^l$: Bandwidth usage of #k application
$B_{i,k}^p$: Processing time of #k application on #i device

図 3　線形計画式のパラメータ

り，(2) のコストが幾ら以内は制約条件の一つとなり，(3)(4) のサーバのリソース上限を超えていないかの制約条件も加わる．応答時間要求により，アプリケーションの応答時間が何秒以内での配置が必要な要求の場合は，(2) に対応する (5) のコストの最小化が目的関数となり，(1) に対応する (6) の応答時間が何秒以内は制約条件の一つとなり，(3)(4) の制約条件も加わる．

(1) 及び (6) はアプリケーション k の応答時間 $R_k$ を計算するための式であり，(1) の場合は目的関数，(6) の場合は制約条件である．(2) 及び (5) はアプリケーション k を動作させるコスト（価格）$P_k$ を計算するための式であり，(2) の場合は制約条件，(5) の場合は目的関数である．(3)(4) は，計算リソース及び通信帯域の上限を設定する制約条件であり，他者が配置したアプリケーション含めて計算され，新規ユーザのアプリケーション配置によるリソース上限の超過を防ぐ．

(1)-(4) 及び，(3)-(6) の線形計画の式を，ネットワークトポロジーや変換アプリケーションタイプ（CPU に対するコスト増と性能増等），ユーザ要求，既配置アプリケーションの異なる条件に対して，GLPK（Gnu Linear Programming Kit）や CPLEX（IBM Decision Optimization）等の線形計画ソルバで解を導出することで，適切なアプリケーション配置を計算できる．適切配置計算後に実際の配置を，複数のユーザに対して，順次行っていくことで，複数のアプリケーションが各ユーザの要求に基づいて配置される．

$$R_k = \sum_{i \in Device} (A_{i,k}^d \cdot B_{i,k}^p) + \sum_{j \in Link} (A_{j,k}^l \cdot \frac{C_k}{B_k^l}) \tag{1}$$

$$\sum_{i \in Device} a_i (\frac{A_{i,k}^d \cdot B_k^d}{C_i^d}) + \sum_{j \in Link} b_j (\frac{A_{j,k}^l \cdot B_k^l}{C_j^l}) \leqq P_k \tag{2}$$

$$\sum_{k \in App} (A_{i,k}^d \cdot B_k^d) \leqq C_i^d \tag{3}$$

$$\sum_{k \in App} (A_{j,k}^l \cdot B_k^l) \leqq C_j^l \tag{4}$$

$$P_k = \sum_{i \in Device} a_i (\frac{A_{i,k}^d \cdot B_k^d}{C_i^d}) + \sum_{j \in Link} b_j (\frac{A_{j,k}^l \cdot B_k^l}{C_j^l}) \tag{5}$$

$$\sum_{i \in Device} (A_{i,k}^d \cdot B_{i,k}^p) + \sum_{j \in Link} (A_{j,k}^l \cdot \frac{C_k}{B_k^l}) \leqq R_k \tag{6}$$

## 4. 評　　価

線形計画手法の定式に基づき，無償ソルバの GLPK を用いて，複数のアプリケーションが適切に配置されていくことを，幾つかの条件を変更して確認する．

### 4.1 評価条件

#### 4.1.1 対象アプリケーション

配置対象のアプリケーションは，多くのユーザが利用すると想定されるフーリエ変換と画像処理とする．

フーリエ変換処理（FFT）は，振動周波数の分析等，IoT でのモニタリングの様々な場面で利用されている．NAS.FT [38] は，FFT 処理のオープンソースアプリケーションの一つである．備え付けのサンプルテストの 2048*2048 サイズの計算を行う．IoT で，デバイスからデータをネットワーク転送するアプリケーションを考えた際に，ネットワークコストを下げるため，デバイス側で FFT 処理等の一次分析をして送ることは想定される．

MRI-Q [39] は，非デカルト空間の 3 次元 MRI 再構成アルゴリズムで使用されるキャリブレーション用のスキャナー構成を表す行列 Q を計算する．IoT 環境では，カメラビデオからの自動監視のために画像処理が必要になることが多く，画像処理の自動オフロードへのニーズはある．MRI-Q は C 言語アプリケーションで，パフォーマンス測定中に 3 次元 MRI 画像処理を実行し，Large の 64*64*64 サイズのサンプルデータを使用して処理時間を測定する．CPU 処理は C 言語で，FPGA 処理は OpenCL に基づき処理される．

著者の GPU，FPGA 自動オフロード技術 [24] [27] により，NAS.FT は GPU で高速化でき，MRI-Q は FPGA で高速化でき，それぞれ，CPU に比べて 5 倍，7 倍の性能がでることが分かっている．

#### 4.1.2 評価手法

アプリケーションを配置するトポロジーは図 3 の様に 3 層で構成され，クラウドレイヤーの拠点数は 5，キャリアエッジレイヤーは 20，ユーザエッジレイヤーは 60，インプットノードは 300 とする．IoT 等のアプリケーションを想定してインプットノードから IoT データ等がユーザエッジに収集され，アプリケーションの特性（応答時間の要求条件等）に応じて，ユーザエッジ，キャリアエッジで分析処理がされたり，クラウドまでデータをあげてから分析処理されたりされる．

分析処理するサーバは全て単一事業者の保持資産であり，サーバやリンクの上限値や価格は，事業者が決める．今回の評価実験では，著者が以下の方針で決めた．各拠点に，クラウドでは，サーバは CPU サーバ 8 台，GPU 16GB RAM のサーバ 4 台，FPGA サーバ 2 台，ネットワークエッジでは，CPU サーバ 4 台，GPU 8GB RAM サーバ 2 台，FPGA サーバ 1 台，ユーザエッジでは，CPU サーバ 2 台，GPU 4GB RAM サーバ 1 台とする．サーバコストについて，CPU，GPU，FPGA サーバについて，サーバの標準的価格として 60 万，120 万，144 万を 1 年で回収するとし，サーバ 1 台の全リソース使用（GPU では 16GB RAM 利用時）の月額は 5 万，10 万，12 万とした．集約

効果のため，ネットワークエッジ，ユーザエッジは割高になるとして，クラウドの 1.25 倍，1.5 倍の月額とした．

リンクについては，クラウド-キャリアエッジ間は 100Mbps，キャリアエッジ-ユーザエッジ間は 30Mbps の帯域が確保されている．リンクコストについては，IoT サービス向けの OCN モバイル One フル MVNO の価格等を参考に（データ転送量 500MB まで月 500 円，1GB まで月 800 円等），独自に 100Mbps のリンクは月額 8000 円，30Mbps のリンクは月額 5000 円とした．

アプリケーションが利用するリソースとして，処理時間等は，実際に GPU，FPGA にオフロードした際の値を用いる．NAS.FT は，利用リソース量は GPU 1GB RAM，利用帯域 2Mbps，転送データ量 0.2MB，処理時間 5.8 秒である．MRI-Q は，利用リソース量は FPGA サーバの 10%（FlipFlop，LookUPTable 等の利用数が FPGA の利用リソースとなる），利用帯域 1Mbps，転送データ量 0.15MB，処理時間 2.0 秒である．

パラメータ値を元に，ユーザ要求条件に基づいて 1000 個のアプリケーションを配置する．アプリケーションは，IoT アプリケーションで，インプットノードから生じるデータを分析する想定である．300 のインプットノードから配置依頼をランダムに生じさせる．

配置依頼数として，NAS.FT : MRI-Q=3:1 の割合で 1000 回アプリを配置依頼．

ユーザ要求として，配置依頼する際にアプリ毎に価格条件か応答時間条件が選ばれる．NAS.FT の場合，価格に関しては月 7000 円上限か 8500 円上限か 10000 円上限，応答時間に関しては 6 秒上限か 7 秒条件か 10 秒上限が選択される．MRI-Q の場合，価格に関しては月 12500 円上限か 20000 円上限，応答時間に関しては，4 秒上限か 8 秒上限が選択される．ユーザ要求のバリエーションとして，3 パターンがある．パターン 1：NAS.FT では 6 種のリクエストを 1/6 ずつ，MRI-Q では 4 種のリクエストを 1/4 ずつ選択する．パターン 2：リクエストは最低価格が上限の条件を選択（最初は 7000 円，12500 円）し，空きがない場合は次に安い価格条件とする．パターン 3：リクエストは最低応答時間が上限の条件を選択（最初は 6 秒，4 秒）し，空きがない場合は次に速い応答時間条件とする．

#### 4.1.3　評価ツール

配置は，ソルバ GLPK5.0 を用いてシミュレーション実験により行う．規模のあるネットワーク配置の模擬のためツールを用いたシミュレーションになる．実利用の際は，アプリケーションのオフロード依頼が来たら，検証環境を用いた繰返し性能試験でオフロードパターンを作成し，検証環境での性能試験結果に基づいて適切なリソース量を決め，ユーザ要望に応じて GLPK 等用いて適切な配置を定め，実際にデプロイした際の正常確認試験や性能試験を自動で行い，その結果と価格をユーザに提示して，ユーザ判断後利用を開始する．

### 4.2　結　　果

図 4 は，平均応答時間とアプリケーション配置数を 3 パターンに対して取ったグラフである．

パターン 2 ではクラウドから順に，パターン 3 ではエッジから順に埋まっていくことが確認できた．パターン 1 では，多様な依頼が来た際に，ユーザ要求条件を満たして配置される．

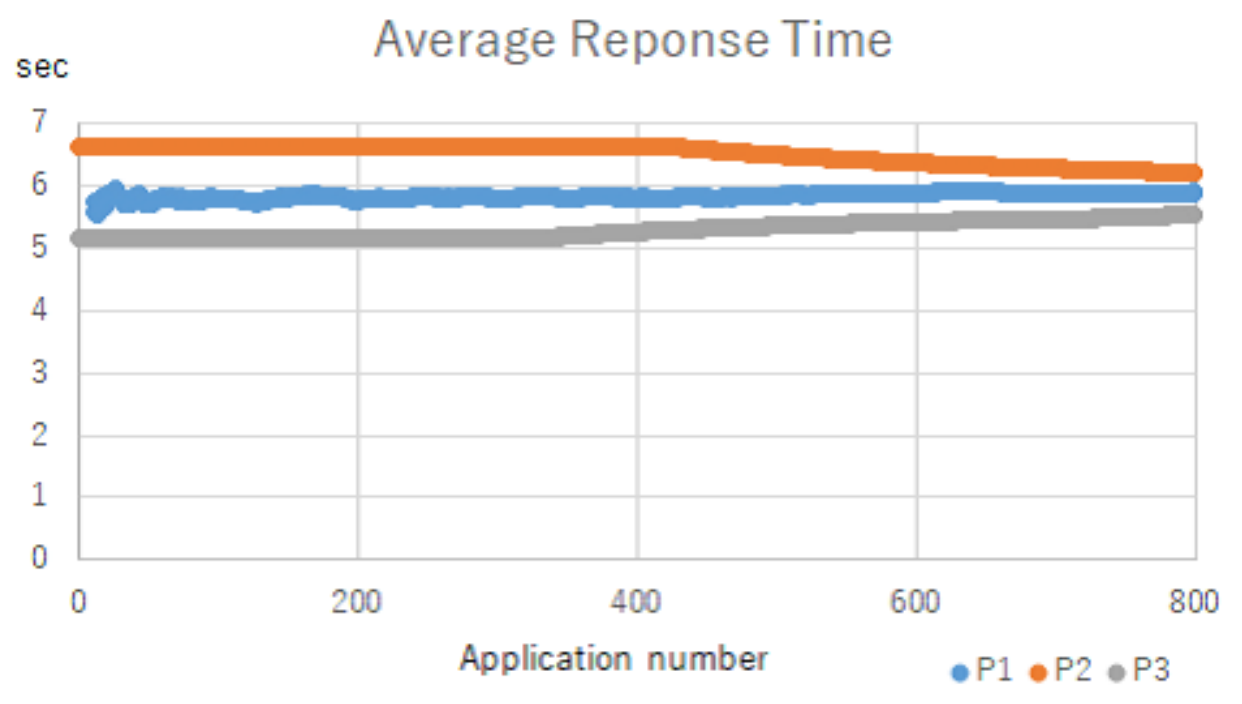

図 4　平均応答時間のアプリケーション配置数変化

図 4 に関して，パターン 2 では 400 配置位までは全てクラウドに配置され平均応答時間は最遅のままであるが，クラウドが埋まると段々下がっていくことになる．パターン 3 では NAS.FT はユーザエッジから，MRI-Q はキャリアエッジから配置されるため平均応答時間は最短であったが，数が増えるとクラウドにも配置されるため平均応答時間は遅くなる．パターン 1 に関しては，平均応答時間は，パターン 2 や 3 の中間であり，ユーザ要求に応じて配置されるため，最初はクラウドに全て入るパターン 2 に比べて平均応答時間は適切に低減されている．

## 5.　ま　と　め

本稿では，私が提案している，ソフトウェアを配置先環境に合わせて自動適応させ GPU，FPGA，メニーコア CPU 等を適切に利用して，アプリケーションを高性能に運用するための環境適応ソフトウェアの要素として，GPU 等に自動オフロードした際に，ユーザのコスト要求，応答時間要求に答えるため，アプリケーション配置の適正化手法を提案した．

GPU 等のデバイスで処理できるよう，プログラムを変換し，アサインするリソース量が定まった後に提案方式は動作する．提案方式は，まず，プログラム変換する際に検証環境で行った性能試験のデータから，アプリケーションの利用データ容量，計算リソース量，帯域，処理時間を設定する．変換アプリケーション毎に設定される値と，事前に設定されるサーバやリンクのコスト等の値から，線形計画式に基づき，アプリケーションの適切な配置が計算される．アプリケーション配置の際は，ユーザが指定する価格や応答時間のリクエストに基づき，一方が制約条件にもう一方が目的関数となる．線形計画ソルバにより適切な配置が計算され，提案方式は計算された場所にリソースを配置した際の，価格等をユーザに提示し，ユーザ承諾後に利用が開始される．

GPU，FPGA に自動オフロードしたアプリケーションに対して，ユーザのリクエストする価格条件や応答時間条件，アプリケーションの配置数等を変更して，提案方式で適正配置を計算し，方式の有効性を確認した．今後は，適正配置をアプリ

ケーション利用開始の最初にだけ計算するのでなく，運用中でもより適正な配置や変換オフロードパターンがある際に，再構成する運用中再構成の検討を行う．

## 文　　献